\documentclass[aps,twocolumn,amsmath,floatfix]{revtex4-1}
\usepackage{amssymb}
\usepackage{graphicx}
\usepackage{array}
\usepackage{hhline}
\usepackage{longtable}
\usepackage{bm}

\begin{document}

\newcommand{\rsp}[1]{\hspace{-0.15em}#1\hspace{-0.15em}}

\title{Intrinsic curvature determines the crinkled edges of ``crenellated disks''}

\author{C. Nadir Kaplan$^{1, 2}$, Thomas Gibaud$^{1, 3}$, and Robert B. Meyer$^{1, 4}$}
\affiliation{$^{1}$The Martin Fisher School of Physics, Brandeis University, Waltham, Massachusetts 02454, USA
\\$^{2}$School of Engineering and Applied Sciences, Harvard University, Cambridge, Massachusetts 02138, USA
\\$^{3}$Laboratoire de Physique, ENS Lyon, Universite´ de Lyon I, CNRS/UMR 5672, 46 Allée d'Italie, 69007 Lyon, France
\\$^{4}$Department of Physics, Wellesley College, Wellesley, Massachusetts 02481, USA}

\date{\today}

\begin{abstract}
Elastic curvature constants determine many structural and functional properties of fluid membranes. Methods to measure the mean curvature modulus have proved to be robust. In contrast, Gaussian curvature is an intrinsic property of a surface. Thus, measuring the relevant modulus $\bar{k}$ in fluid membranes remains a challenging task. Inspired from colloidal ``crenellated disks'' observed in a model system composed of hard rods, we propose a concise relation between the two curvature moduli and the parameters associated with the free, crinkled edges. Our approach offers a straightforward way to determine $\bar{k}$ of these reconfigurable membranes, where various complex topologies can be nanosculpted. Further, we reveal the structure and stability of the ``crenellated disks.'' 
\end{abstract}
\pacs{pacs} 
\maketitle

\section{Introduction}

A significant achievement towards a continuum description of two dimensional (2D) fluid membranes was made by Helfrich, when he proposed a free energy expansion in terms of mean (extrinsic) and Gaussian (intrinsic) curvatures, the two invariants of the curvature tensor of a surface~\cite{Helfrich1}. 
It follows that many physical and functional characteristics of membranes depend on the associated elastic curvature constants, such as diverse vesicle shapes in fluid membranes and specifically red blood cells~\cite{Helfrich2, Lipowsky, Helfrich3}, fission and fusion processes~\cite{Joanny, Siegel}, formation of caveolae and buds in lipid bilayers or cell walls~\cite{Rafts, Buds}, cell division, and many more. Microscopically, these elastic curvature constants are determined by lateral pressure profiles across a membrane~\cite{Szleifer}, which are obtained through theoretical models or extensive computer simulations~\cite{Deserno, Deserno2, 3DPressure, Sonne}. As for experiments, methods to measure mean curvature modulus have become reliable, such as analysis of thermal undulations along the membrane normal~\cite{Barry2, Helfrich4} or vesicle aspiration and tube pulling techniques~\cite{Baumgart}. 
On the other hand, the Gaussian curvature modulus can only be measured when \textit{e.g.} catenoidal pores nucleate in a membrane~\cite{Siegel}, or by the characterization of the boundary curve enclosing the membrane surface. These measurements utilize the well-known Gauss-Bonnet theorem~\cite{Stoker, Millman}, which relates the boundary geometry with the total Gaussian curvature of the area it surrounds. However, these transformations of membrane conformation are in general difficult to observe and control, posing a problem for the measurement of the intrinsic curvature modulus.

In this paper we develop a theoretical model which focuses on the boundary properties of ``crenellated disks" observed in a model system of colloidal monolayers assembled from chiral, rod-like $fd$ viruses~\cite{Barry2, gibaud}. We simultaneously minimize the total free energy, which is based on the Helfrich framework, both with respect to the surface and the boundary deformations of a crenellated disk. Our approach reveals a simple analytical relation between the two elastic curvature moduli and the parameters characterizing the crinkled edges. This relation serves as a powerful tool to indirectly measure the Gaussian curvature modulus of these membranes. Then we determine the equilibrium structure of the crenellated disks. Finally, we investigate their thermodynamic stability with respect to the perfectly flat monolayers, which are widely observed in the global phase diagram of $fd$-viruses mixed with a depleting agent.

The present work is organized as follows: In the next section we explain the hypothesized structure of the crenellated disks. In Sec.~\ref{sec:Theory} we present a detailed description of our model. In Sec.~\ref{sec:Stability} we discuss our theoretical results on the structure of the crenellated disks and the theoretical phase diagram. Concluding remarks are offered in the final section of the paper.

\section{Structure of a ``crenellated disk''}
\label{sec:Structure}

\begin{figure}
\centering
\includegraphics[width=1\columnwidth]{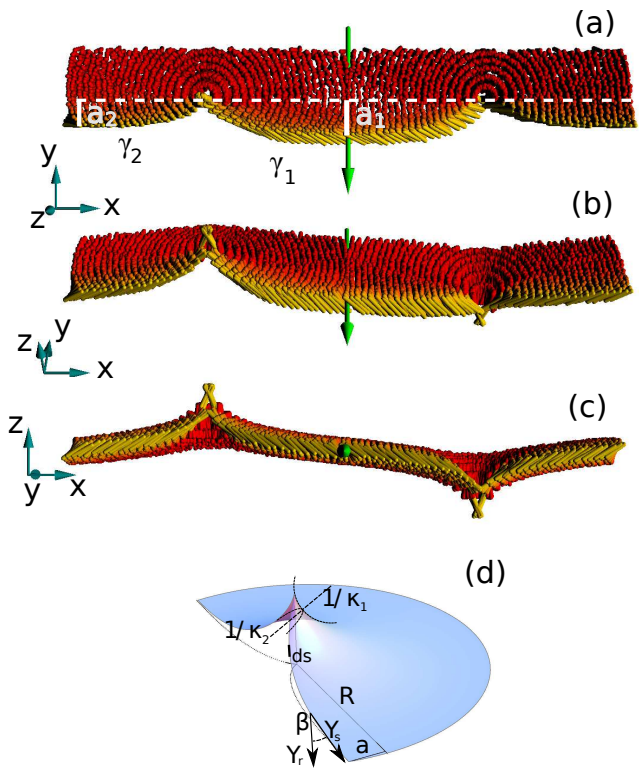}
\caption{\textbf{Schematics of the crinkled edges surrounding a crenellated disk.} (a) Top, (b) top-front, (c) front views of the orientation of rods around the cusps. Adjacent cusps alternate above and below the monolayer plane. In (a) the line tension $\gamma_i$ and in-plane protrusion amplitude $a_i$ of each free boundary are also shown. The green arrow is the axis of reflection. Away from the cusps the bulk of the monolayer is planar. For illustration purposes the concentration and the aspect ratio of the viruses have been reduced.  (d) Schematics of an isolated cusp. For definitions, see text.}
\label{fig:cookiesketch}
\end{figure}

Experimentally, mixtures of monodisperse rods with opposite chirality are prepared from two types of viruses: a left-handed wild-type $fd$ virus ($fd$ $wt$) and a right-handed mutant $fd$ virus, dubbed $fd$ Y21M. These mixtures were observed to form a nematic phase at a certain mixing ratio, and a cholesteric ($Ch$) phase with a reduced overall strength of handedness otherwise~\cite{Barry3}. In the presence of a depleting agent, rod mixtures which are either achiral or have a low overall chirality assemble into the crenellated disks. These unusual structures are essentially flat monolayers surrounded by an array of local, three-dimensional (3D) cusp defects (Figs.~\ref{fig:cookiesketch}(a)-(c))~\cite{gibaud2}. In the membrane interior, the rods are perfectly aligned parallel to the layer normal, resembling a single layer of a conventional chiral smectic-$A$ (Sm-$A^\ast$) phase~\cite{Barry}. Molecular twist deformations penetrating at the edge are either clockwise or anticlockwise, and surround the monolayer 
in the form of a narrow $Ch$ band. The cusps at the boundaries are the termination points of these alternating left- and right-handed $Ch$ bands. Note that phase separation between the two types of viruses is observed neither in the crenellated disks nor in bulk mixtures. Instead of forming defects on the plane, these $Ch$ regions escape into a third dimension in order to reduce the orientational mismatch of rods where the edges meet (Fig.~\ref{fig:cookiesketch}(c)). Because molecular tilt changes direction at each of these reversal points, adjacent cusps must alternate above and below the plane of the otherwise flat disk. Towards the defect peak, each of these $Ch$ sections acquires a macroscopic handedness opposite to the direction of molecular tilt, \textit{i.e.} a $Ch$ band becomes a left-handed space curve when molecular twist deformations are right handed, and vice versa (see the yellow bands in Figs.~\ref{fig:cookiesketch}(a) and (b)). Likewise, in a previous work rods with given chirality were found 
to form twisted ribbons (minimal surfaces to a double helix) of opposite handedness as a result of molecular twist deformations~\cite{Kaplan}. To maintain a continuous structure between two adjacent cusps, there lies an axis of reflection, as shown in Fig.~\ref{fig:cookiesketch} (green arrows).

In contrast to lipid monolayers and bilayers~\cite{Marsh}, $fd$ viruses tend to build colloidal membranes with negative Gaussian curvature~\cite{gibaud, Kaplan}. In addition to twisted ribbons, which occur in the limit of high chirality and low depletant concentration, the strong effect of negative Gaussian curvature on the Sm-$A^\ast$ membranes is evidenced by two further observations: (1) Disks obtain high edge energies by typically reaching to mesoscopic sizes in diameter, however, they never fold into surfaces without boundaries, such as a sphere (a surface with positive Gaussian curvature), to avoid these free edges, (2) achiral disks stretched by optical forces again form a twisted ribbon~\cite{gibaud}.  Furthermore, a characteristic of a saddle surface is that it exposes a longer edge than if it were flat~\cite{Sheets}, which is indeed observed both in the twisted ribbons and crenellated disks. Thus, we hypothesize that the formation of cusp arrays could only be stabilized with respect to the flat 
monolayers 
by virtue of the negative Gaussian curvature. The saddle characteristics of the surface further increase by the formation of in-plane protrusions (Fig.~\ref{fig:cookiesketch}(a)) at the expense of increasing interfacial energy along the free edges.

\section{Theory}
\label{sec:Theory}

Here, we assume rotational symmetry around each cusp. That is, we consider a surface of revolution with negative Gaussian curvature for an isolated defect, where the peak is located at its rotation axis (Fig.~\ref{fig:cookiesketch}(d)). The variation of the free energy of a generic membrane with respect to the surface deformation has been calculated within the context of chiral lipid membranes~\cite{ZhongCan1, ZhongCan3, Seifert}. To account for the formation of the in-plane protrusions as well, we carry out a minimization both with respect to the deformation of the area and the free boundaries of the membrane. We employ the Helfrich free energy along with relevant interfacial terms, and ignore the variation of the molecular director field to simplify the analysis.  Then the free energy of the Sm-$A^\ast$ membranes is given as follows:
\begin{equation}
\begin{split}
& F=\int f_H dS+ \sum_{i=1,2}\oint ds_i (\gamma_i+k_b \kappa_{s, i}^2-h_i\tau_{g,i})\,,
\\&\text{where}\quad f_H=\frac{k}{2} (2 H)^2+\bar{k} \kappa_G+\sigma\,.
\end{split}
\label{eq:freeE}
\end{equation}
$f_H$ is Helfrich free energy density of surface bending, where $H$ is the mean curvature, $k_G$ is the Gaussian curvature, and $k$, $\bar{k}$ are the corresponding moduli, respectively~\cite{Helfrich1, Helfrich2}. $\sigma$ is the surface tension of the membrane. These terms are multiplied by $dS$, the membrane area element. The interfacial terms, associated with the formation of the free edges, are multiplied by the infinitesimal arc length $ds_i$ (see Fig.~\ref{fig:cookiesketch}(d)). Each cusp is the termination point of two free edges with opposite handedness, hence the sum in front of the closed integral. That is, two edge profiles are allowed to be different. First two interfacial terms are; the line tension $\gamma_i$, and $\kappa_b$, the bending rigidity of the interface. Experimentally, $\gamma_i$ and $\kappa_b$ are measured from the power spectra of interfacial fluctuations on the monolayer plane. The interfacial tension $\gamma_i$ is controlled by the molecular chirality of the viruses, as well as 
the 
depletant concentration\cite{gibaud}. 
The natural length scales of the system, which determine the relaxation of the 3D structure into a 2D membrane, are given by $\xi_i\equiv\sqrt{k_b/\gamma_i}$. The geodesic torsion $\tau_{g, i}$ in the last term of Eq.~\eqref{eq:freeE} is the pseudoscalar describing the macroscopic chirality of the alternating left- and right-handed $Ch$ sections. Its prefactor is $h_i$. 
This term serves as a correction to $\gamma_i$ due to the 3D structure of the cusp, and vanishes away from it along the edge. In Sm-$A^\ast$ membranes, because molecular twist is confined to a narrow region at the interface, we approximate $-h_i\tau_{g, i}$ to an interfacial term by assuming a constant molecular tilt within a narrow band at the edge. This is in contrast with the case of chiral lipid membranes, where the tilt of the chiral molecules is uniform throughout the membrane, leading to a bulk term $-h\tau_{g}$~\cite{ZhongCan2, ZhongCan1, Seifert}.

We evaluate Eq.~\eqref{eq:freeE} for a generic surface of revolution of negative intrinsic curvature without making any further assumptions, except that the local height of the surface $f(r)$ is a single-valued function. In Euclidean three-space a surface of revolution of radius $R$ is represented by the position vector 
\begin{equation}
\begin{split}
\label{eq:surface} 
\mathbf{Y}(r, \varphi)&=\{r\cos\varphi, r\sin\varphi, f(r)\}\,,
\\&\text{where}
\quad r\leq R\,,\quad -\phi_1(r)\leq\varphi\leq\pi+\phi_2(r)\,, 
\\& \text{and} \quad\phi_1(r)\,,\phi_2(r)>0\,.
\end{split}
\end{equation}
We define a set of orthogonal coordinate vectors $\{\mathbf{Y}_r\,, \mathbf{Y_\varphi}\,, \hat{\mathbf{N}} \}$ on the membrane, where $\hat{\mathbf{N}}$ is the local unit normal and $\mathbf{Y}_r\,, \mathbf{Y_\varphi}$ are the tangent vectors to the surface in radial and azimuthal directions, respectively. The two invariants of the surface, the first- and second-fundamental forms, can be defined in terms of an angle $\alpha(r)<0$ and its derivatives, which satisfies $\tan{\alpha(r)}\equiv f'$ (primes denote derivatives with respect to $r$). The two principal curvatures via these invariants are obtained as $\kappa_1=\cos\alpha\alpha'$ and $\kappa_2=\sin\alpha/r$ (Fig.~\ref{fig:cookiesketch}(d)). Then, the mean and Gaussian curvatures are calculated by the relations $H=(\kappa_1+\kappa_2)/2$ and $\kappa_G=\kappa_1\kappa_2$. The area element in terms of $\alpha$ is given by $dS=r dr d\phi/\cos\alpha$.

The surface of revolution is terminated by two space curves confined to the membrane, corresponding to two free edges meeting at the peak. Each individual curve is represented by  the position vector $\mathbf{Y}_i(s)=\mathbf{Y}_i(r(s), \phi_i(s))$ on the surface, where $s$ is the coordinate along the arc length of the curve~\cite{Stoker}. The curvature of these curves are obtained in terms of $\alpha$, $\beta_i(r)$, and their derivatives, where $\beta_i(r)<0$ is the angle between the tangent vector $\mathbf{Y}_{s, i}$ to the $i$-th curve and $\mathbf{Y}_r$ (Fig.~\ref{fig:cookiesketch}(d)). For $\alpha$, $\beta_i$, and $\phi_i$ the relation $\tan\beta_i=r\cos\alpha\phi_i'$ always holds. This relation serves as an additional constraint to be added to Eq.~\eqref{eq:freeE} with an associated Lagrange multiplier $\lambda_i(r)$, arising from the choice of parametrization. The total curvature squared of the free edges is given by $\kappa_{s, i}^2=\kappa_{g, i}^2+\kappa_{n, i}^2$, where $\kappa_{g, i}=\frac{1}{r}\cos
\alpha\left(r\sin\beta_i\right)'$ is the geodesic curvature and $\kappa_{n, i}=\kappa_1\cos^2\beta_i+\kappa_2\sin^2\beta_i$ is the normal curvature of the lines~\cite{Millman}. The geodesic torsion of each curve is found as $\tau_{g,i}=\sin\beta_i\cos\beta_i(\kappa_2-\kappa_1)$. The line elements in Eq.~\eqref{eq:freeE} are given by $ds_i=dr/(\cos\alpha\cos\beta_i)$ (Fig.~\ref{fig:cookiesketch}(d)).

We perform a mean-field theory by simultaneously minimizing Eq.~\eqref{eq:freeE} with respect to the deformation of the surface and the free edges. This is done by analyzing the variations of the free energy $\frac{\delta F}{\delta g_j}$ as a boundary-value problem, where $g_j=\{\alpha, \beta_1, \beta_2, \phi_1, \phi_2\}$ ~\cite{supplemental}. Far from the rotation axis the minimum of the free energy $\frac{\delta F}{\delta g_j}=0$ is achieved using the following boundary conditions: 
The surface becomes a flat membrane where $f(r)$ and all its derivatives vanish. Hence, $\kappa_1$, $\kappa_2$, and $\kappa_{n, i}$ become zero. Furthermore, due to two torque-free boundary conditions the curvature $\kappa_{s, i}$ of the free edges ideally vanish, resulting in a straight line in the flat membrane limit ($\kappa_{g, i}=0$, $\beta_i=-\phi_i$ in this limit), and $\lambda_i(r)=0$~\cite{supplemental}. Plugging in these boundary conditions to the set of $\frac{\delta F}{\delta g_j}$, we arrive at the relation

\begin{equation}
k+\bar{k}=\gamma_1 A_1=\gamma_2 A_2\,,
\label{eq:kbarvsamplitude} 
\end{equation}
where $A_i$ is the maximum in-plane protrusion amplitude of the $i$-th edge (that is, $a_i\bigl|_{R\gg\xi_i}=A_i$, see Fig.~\ref{fig:cookiesketch}(a)). Eq.~\eqref{eq:kbarvsamplitude} is a natural consequence of the Gauss-Bonnet theorem, which states that the total Gaussian curvature of an area is equivalent to the total amount of the geodesic curvature $\kappa_{g, i}$ and jump angles along its boundaries~\cite{Stoker, Millman}. This ensures a practical determination of the Gaussian curvature modulus $\bar{k}$, provided that measuring the line tensions $\gamma_i$, the mean curvature modulus $k$, and the amplitudes $A_i$ is relatively easy in experiments. Furthermore, it implies an elegant connection between $A_1$ and $A_2$, specified solely by $\gamma_1$ and $\gamma_2$, or vice versa. 

\section{Shape and stability of crenellated disks} 
\label{sec:Stability}

We apply Eq.~\eqref{eq:freeE} to the array of cusps surrounding Sm-$A^\ast$ monolayers composed of virus mixtures in achiral and slightly chiral regimes.  In the latter, one of the $Ch$ bands (\textit{e.g.} associated with $\gamma_2$) has opposite chirality to the overall handedness of the mixture. Thus, it chooses an erroneous tilt direction, whereas the other $Ch$ band (with $\gamma_1$) tilts the right way. Eventually, $\gamma_2 >\gamma_1$, and the edge of incompatible chirality becomes shorter (Fig.~\ref{fig:cookiesketch}(a)). This results in a long-range attractive force between two cusps. On the other hand, when two defects are sufficiently close, a short-range repulsive interaction arises primarily from the edge curvature energy. Hence, frustration emerges between line and curvature energies of the free edges, and one should expect a finite equilibrium distance $d_0 (=2R_0)$ between two adjacent peaks. Here $R_0$ is the equilibrium radius of a single defect. For convenience, we denote the edge with 
lower line tension $\gamma_1$, and the edge with higher line tension $\gamma_2$ in the following. Other edge-specific parameters also follow this convention.

In the membranes composed of a single type of virus, the dependence of the line tension on the chirality is reported in Ref.~\cite{gibaud}. In our current theoretical analysis, we model the effect of the overall mixture chirality by $\delta\equiv\gamma_2/\gamma_1$, where $\delta\in\left[1, 1.5\right]$. Furthermore we set $h=0$, since the geodesic torsion has a negligible effect both on the overall shape and the free energy of the crenellated disks. $\kappa_b$ is taken to be $\sim100 k_B T\mu m$~\cite{gibaud}, and its dependence on the mixture chirality is ignored. The lower line tension is measured as $\gamma_1\sim500 k_B T/\mu m$~\cite{gibaud}. In reality, all experimental parameters depend on the chirality of the mixture, which will be published elsewhere along with complementary theoretical results~\cite{gibaud2}.

\begin{figure}
\centering
\includegraphics[width=1\columnwidth]{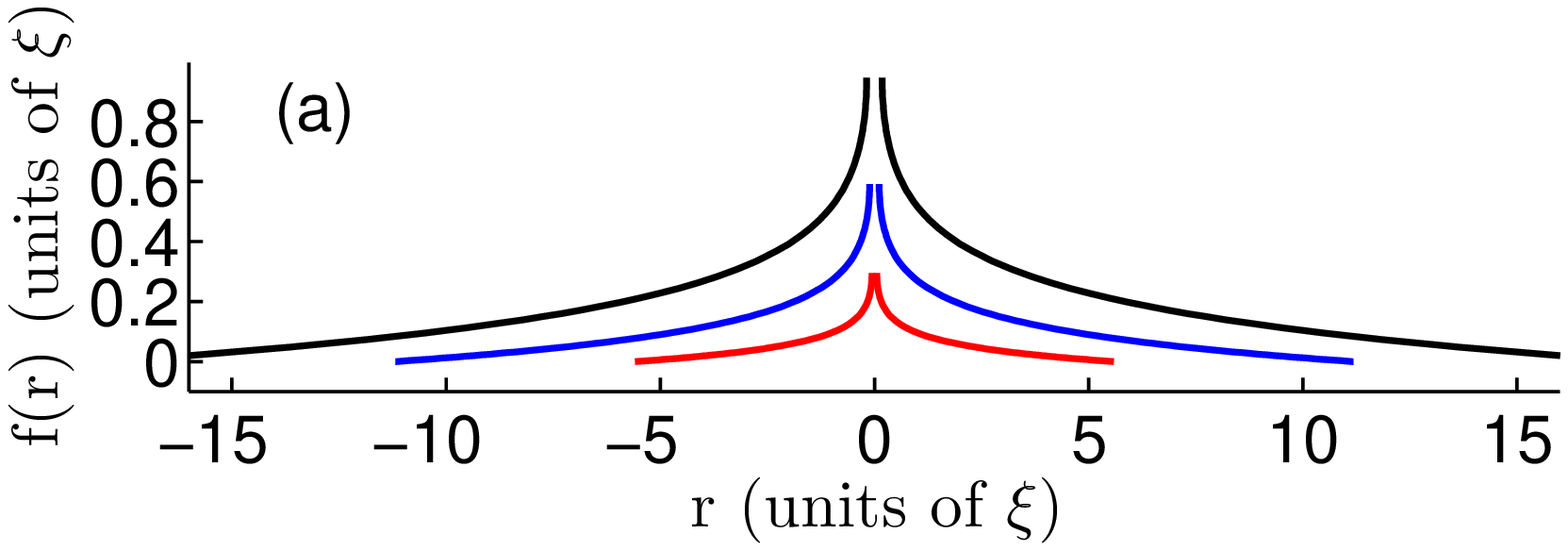}
\includegraphics[width=1\columnwidth]{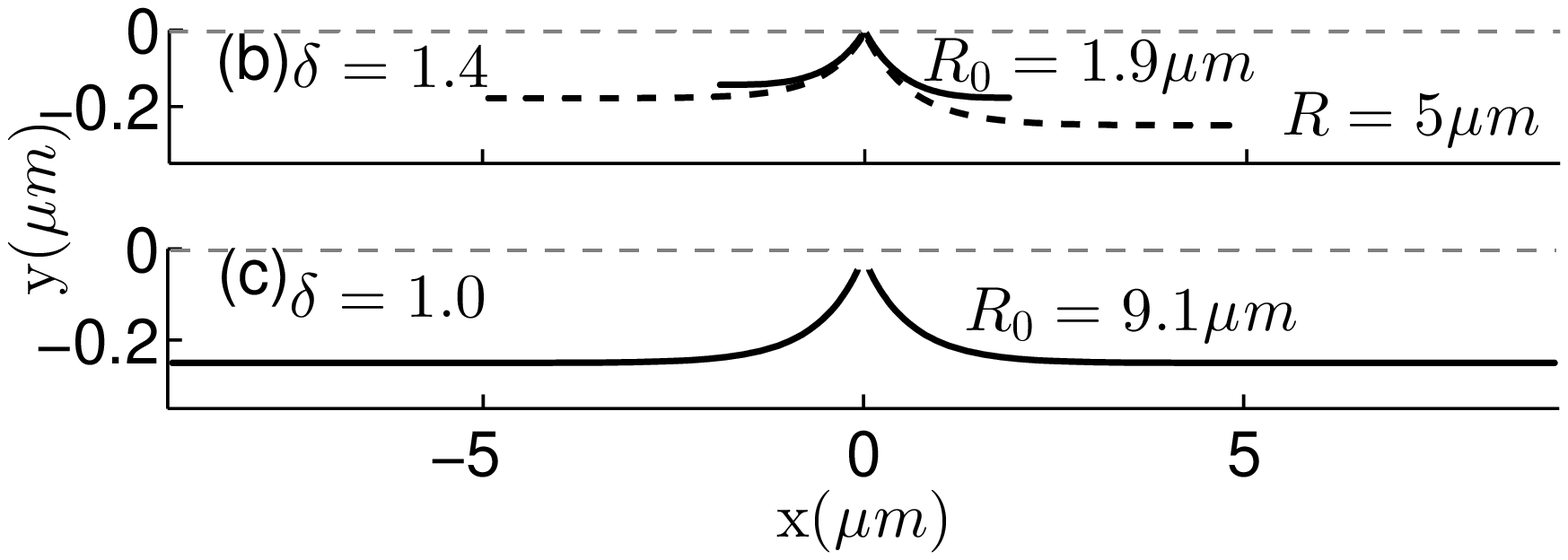}
\caption{\textbf{The profiles of the crenellated disks.} (a) The height $f(r)$ as a function of $r$. The cusp height $f(\epsilon)$ grows with increasing $R$. For $\bar{k}=125k_BT$ the in-plane protrusion profiles are given in cartesian coordinates $y$ versus $x$, corresponding to (b) $\delta=1.4$ (chiral) and (c) $\delta=1$ (achiral). The full curves are equilibrium in-plane protrusions, whereas the black dashed curve in (b) is a profile of a non-equilibrium crenellated disk. The asymmetry due to $\delta\neq 1$ of the edges is obvious in (b). $R_0$ is long for achiral mixtures in (c). Eq.~\eqref{eq:kbarvsamplitude2} holds for sufficiently long edges. }
\label{fig:structure}
\end{figure}

With respect to a perfectly flat monolayer, the formation of cusp arrays is expected to be stabilized by negative Gaussian curvature. Since Sm-$A^\ast$ monolayers have free boundaries and no spontaneous curvature,  the Laplace pressure on the membrane should vanish. Thus, we consider a minimal surface ($H=0$). The only nonplanar minimal surface of rotation is catenoid, where $f(r)$ satisfies
\begin{equation}
\label{eq:catenoid} 
r=\epsilon\cosh\left(\frac{f(r)-f(\epsilon)}{\epsilon}\right)\,,
\end{equation}
as given in Ref.~\cite{Nitsche}. Here $\epsilon\equiv-R\sin\alpha(R)$ is the neck radius of the catenoid ($r\geq\epsilon$), and $f(\epsilon)=-\epsilon\log\left(\frac{\epsilon}{R\left[1+\cos\alpha(R)\right]}\right)$ is the cusp height (Fig.~\ref{fig:structure}(a)). We choose $\alpha(R)$ to be as small as $-10^{-2}$ to suppress any nonanalyticity at $R$ on the surface. The boundary conditions at the cusp are $\alpha=-\pi/2$ and $\beta_i=0$~\cite{supplemental}. While evaluating Eq.~\eqref{eq:freeE}, we ignore the effect of variations $\delta F/\delta\phi_i$ in the regime $a_i\ll R$. Namely, the structure of an isolated cusp defect is mainly governed by $\delta F/\delta \beta_i$ and $f(r)$. 

\begin{figure}
\centering
\includegraphics[scale=0.3]{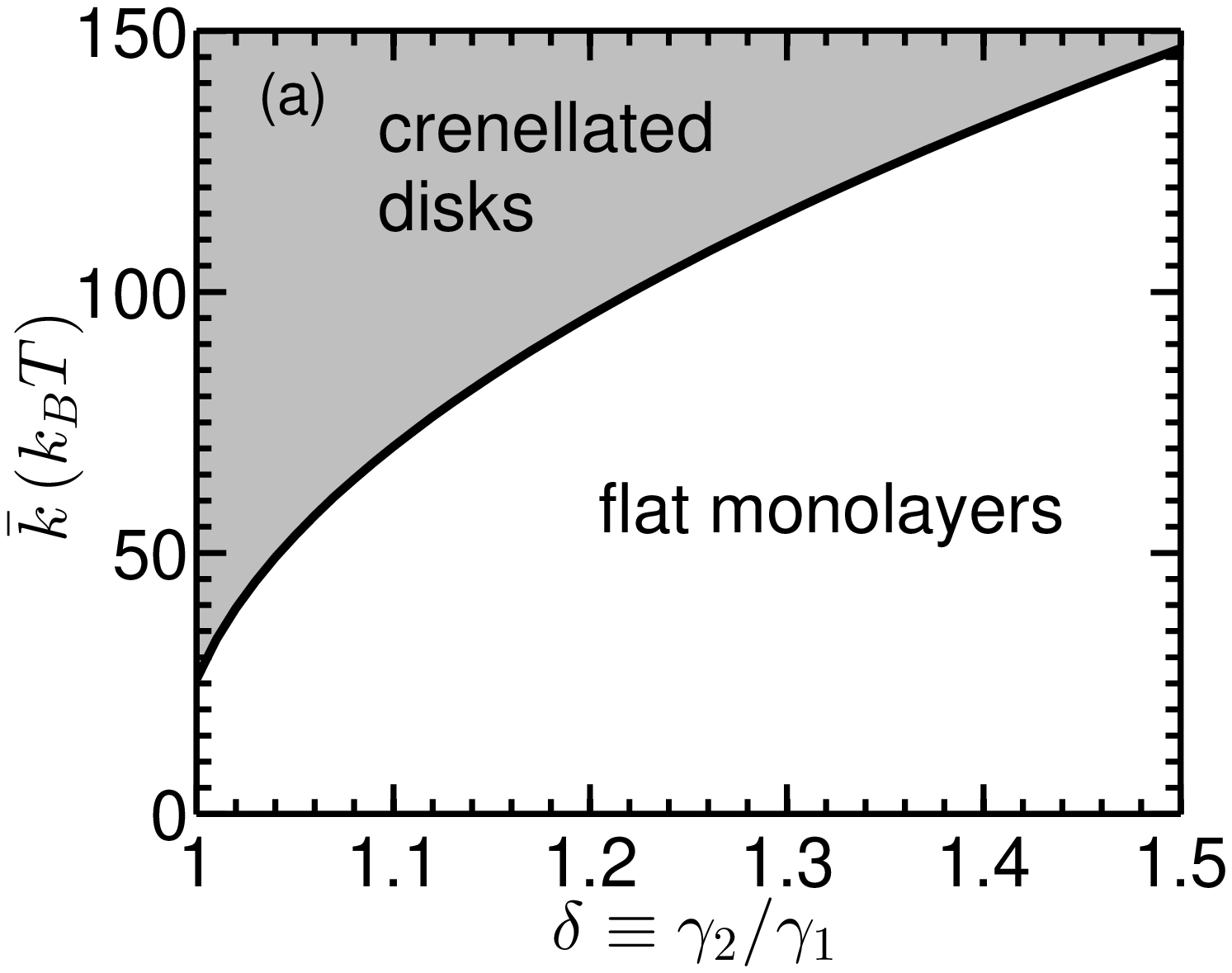}
\includegraphics[width=1\columnwidth]{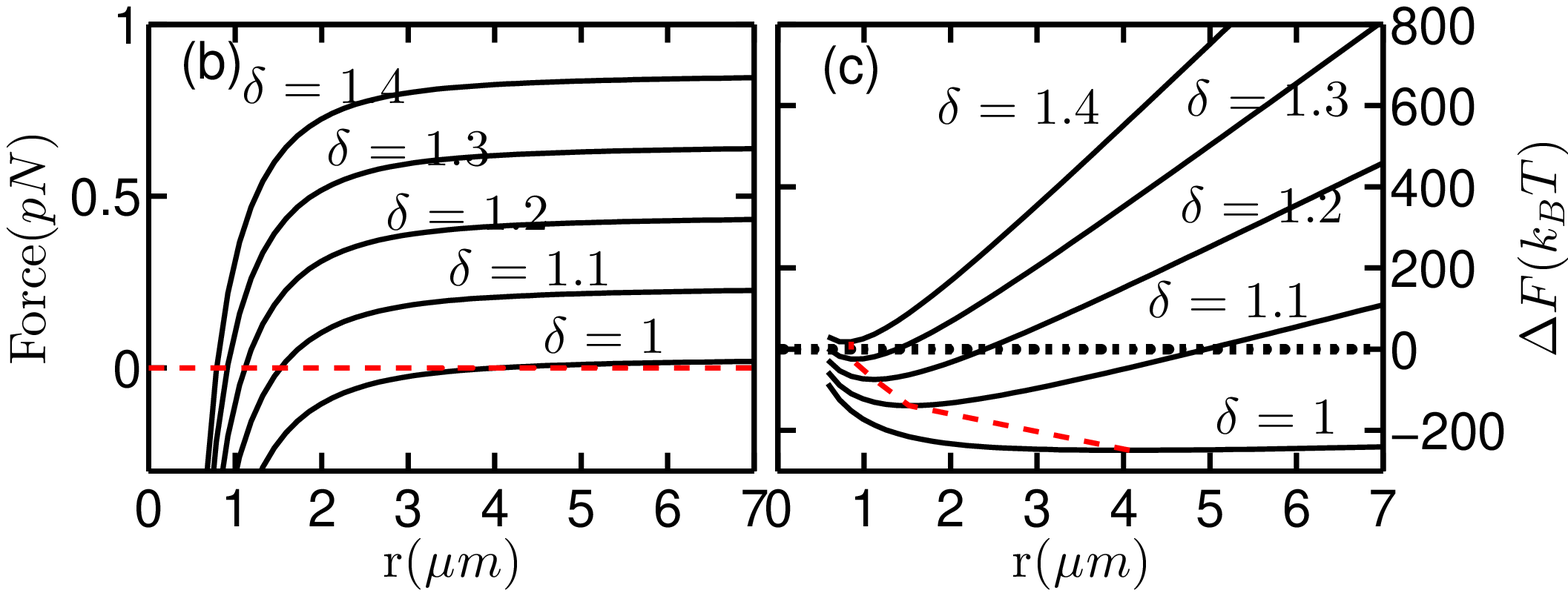}
\caption{ \textbf{The thermodynamic stability of the crenellated disks.} (a) For a minimal surface, phase diagram for the crenellated disks (gray-shaded region) and flat monolayers in $\bar{k}$ versus $\delta\equiv\gamma_2/\gamma_1$. The black full curve is the first order phase boundary. The achiral region is at $\delta=1$. (b) For $\bar{k}=125k_BT$, the force curves ($\equiv\left|-\partial \Delta F/\partial r\right|$) as a function of $r$.  (c) For $\bar{k}=125k_BT$, $\Delta F$ as a function of $r$.  When the minima (red dashed curve) of the curves are below $\Delta F=0$ (black dotted line), then the crenellated disks are stable with respect to the flat monolayers. Otherwise, they are metastable. In (b) and (c), from bottom to top, curves are drawn for increasing $\delta$ with $0.1$ increments, starting from achiral crenellated disks ($\delta=1$).}
\label{fig:phasediagram}
\end{figure} 

The phase diagram for the crenellated disks (gray-shaded area) and flat monolayers as functions of $\bar{k}$ and $\delta$ is displayed in Fig.~\ref{fig:phasediagram}(a). To obtain the phase diagram we calculated the difference $\Delta F$ between the free energy of a single defect (Eq.~\eqref{eq:freeE}) and the free energy $F_{flat}$ of a flat monolayer of an equal area $S$, where $F_{flat}=\sigma S+ 2\gamma_1 R$. 
Both for achiral ($\delta=1$) and chiral mixtures ($\delta>1$), we find a first-order phase transition between the crenellated disks and flat membranes, as $\bar{k}$ is varied. $R_0$ for two adjacent cusps is finite at the phase boundary (see Fig.~\ref{fig:phasediagram}(c)). In reality, because $\gamma_i$ and $\kappa_b$ are different for $fd$ $wt$-rich and $fd$ Y21M-rich mixtures, the phase diagram should exhibit asymmetry between the left- and right-handed samples. However, since we model the effect of chirality solely by $\delta$, Fig.~\ref{fig:phasediagram}(a) does not distinguish between the mixtures of opposite handedness. $\bar{k}$ at the phase transition increases when $R_0$ decreases. Essentially, if $R_0$ is small, then there is less energy gain from the Gaussian curvature due to lower $a_i$ (see Fig.~\ref{fig:structure}(b)). Thus, the Gaussian curvature leads to an additional repulsive force at low distances. The widest region of stability is found in achiral rod mixtures, as also observed in 
experiments. In the \textit{flat membrane phase} shown in Fig.~\ref{fig:phasediagram}(a), the crenellated disks are always found to be metastable, in qualitative agreement with experiments. 

In the absence of the mean curvature it is straightforward that Eq.~\eqref{eq:kbarvsamplitude} simplifies to
\begin{equation}
\bar{k}=\gamma_1 A_1=\gamma_2 A_2\,.
\label{eq:kbarvsamplitude2} 
\end{equation}
When $\delta$ is sufficiently low, $R_0$ is big enough that the equilibrium protrusion amplitude $a_i$ reaches $A_i$, satisfying Eq.~\eqref{eq:kbarvsamplitude2} (see Fig.~\ref{fig:structure}(c)). We find good qualitative agreement between the orders of magnitude of  $A_i$ and the experimental measurements~\cite{gibaud2}. Note that $\bar{k}$ is a material parameter and is expected to be a function of the mixture composition. The measurements of $\bar{k}$ using Eq.~\eqref{eq:kbarvsamplitude2} will be reported elsewhere. The value of $\bar{k}$ estimated in Ref.~\cite{Kaplan} for the Sm-$A^\ast$ membranes was $\bar{k}\sim 10k_B T$, which was too low to explain the case of no spherical vesical formation. In fact $\bar{k}$ for $fd$ virus samples must also be much bigger than the moduli measured for some lipid monolayers~\cite{Marsh, Lipid1, Lipid2} (again, $|\bar{k}|\sim 10k_B T$ for lipid membranes). A geometry with a positive Gaussian curvature always possesses a positive mean curvature, which naturally holds 
for these lipid 
monolayers. Hence, in principle with a lower $|\bar{k}|$ compared to the Sm-$A^\ast$ monolayers, lipid monolayers could stabilize their observed structures.

The strength of the attractive force due to $\gamma_2$ depends primarily on $\delta$. This is evidenced by the in-plane protrusion profiles in Figs.~\ref{fig:structure}(b)--(c), as well as the force curves in Fig.~\ref{fig:phasediagram}(b). When $\delta$ is high, there is a clear asymmetry between two protrusions ($a_1>a_2$), and the attractive force due to $\gamma_2$ is strong. On the other hand, when $\delta$ gets lower, the attractive force becomes weaker as $\gamma_1$ and $\gamma_2$ become comparable to each other, resulting in a bigger $R_0$ (see Fig.~\ref{fig:structure}(c)). In the achiral regime ($\delta=1$) a very weak force on the order of $10^{-4}\gamma_i$ survives, since lower $a_i$ also reduces the edge length between close cusps. In this case, the distance between all adjacent cusps should be equal, as $\gamma_1=\gamma_2$.  As stated above, due to its repulsive nature at low $R$, the negative Gaussian curvature forces $R_0$ to be bigger. A simpler model ignoring the 3D structure of the defects 
and accounting only for the line and curvature energies of the edges consistently resulted in smaller $R_0$ values than the results presented here. This 2D model only gives metastable states. Thereby we deduce that the Gaussian curvature not only stabilizes the crenellated disks, but also ensures an accurate determination of their structural properties.

\section{Conclusion}
\label{sec:Conclusion}

Crenellated disks are monolayers of aligned rods which are decorated with an array of cusps alternating above and below the plane of the disk. At these cusps the molecular twist changes handedness. Previously, Tu and Pelcovits utilized a phenomenological elastic theory of the crenellated disks to calculate the director pattern of the rods on a prescribed surface with variable geometric parameters~\cite{HaoTu}. Our work, rather, puts emphasis on the calculation of the detailed optimum geometry which minimizes the Helfrich free energy. In our analysis we have ignored the variation of the molecular director field. The effect of the director deformations, however, is still accounted for by the interfacial terms in Eq.~\eqref{eq:freeE}, albeit crudely. The constants $\kappa_b$ and $\gamma_i$, which are measured in experiments, are directly related with the liquid crystalline distortions at the edge of the monolayer~\cite{gibaud}. Therefore, the theoretical framework presented here captures the qualitative essence 
of the experimental results while maintaining conciseness. The primary outcome of this model applied to a single, isolated cusp is the connection between the elastic curvature moduli and the quantities specifying the free 
boundaries, as given in Eqs.~\eqref{eq:kbarvsamplitude} and~\eqref{eq:kbarvsamplitude2}. In particular, the relation $\bar{k}=\gamma_i A_i$ provides a novel recipe for the indirect determination of the Gaussian curvature modulus $\bar{k}$ of the rod mixtures in question.

\begin{acknowledgements}
We thank Z. Dogic, R.~A. Pelcovits, P. Sharma, H. Tu, A. Ward, and M.~J. Zakhary for fruitful discussions. This work was supported by the NSF through MRSEC Grant No. 0820492.
\end{acknowledgements}

\end{document}


\newcommand{\rsp}[1]{\hspace{-0.15em}#1\hspace{-0.15em}}

\title{Supplementary Information to ``Intrinsic curvature determines the crinkled edges of ``crenellated disks''.''}

\author{C. Nadir Kaplan$^{1, 2}$, Thomas Gibaud$^{1, 3}$, and Robert B. Meyer$^{1, 4}$}
\affiliation{$^{1}$The Martin Fisher School of Physics, Brandeis University, Waltham, Massachusetts 02454, USA
\\$^{2}$School of Engineering and Applied Sciences, Harvard University, Cambridge, Massachusetts 02138, USA
\\$^{3}$Laboratoire de Physique, ENS Lyon, Universite´ de Lyon I, CNRS/UMR 5672, 46 Allée d'Italie, 69007 Lyon, France
\\$^{4}$Department of Physics, Wellesley College, Wellesley, Massachusetts 02481, USA}

\date{\today}
\maketitle

\subsection{Total free energy}

For a single surface of revolution terminated by two space curves, in terms of the angles $\alpha$, $\beta_i$, and $\phi_i$, geodesic curvatures $\kappa_{g, i}$ and normal curvatures $\kappa_{n, i}$ of the free edges, as well as the principal curvatures $\kappa_1$ and $\kappa_2$ of the surface (defined in the main text), the total free energy is given by

\begin{equation}
\label{eq:Ftotal}
F=F_H+F_{edge_1}+F_{edge_2}+F_{\lambda_1}+F_{\lambda_2}\,, 
\end{equation}
where 
\begin{equation}
\label{eq:FHelfrich}
\begin{split}
F_H=&\int^R_\epsilon dr (\pi+\phi_1+\phi_2) \left[\frac{\sigma r}{\cos\alpha}+\frac{k}{2r}\sin\alpha\tan\alpha\right.\\&\left.+\frac{k}{2} r\cos\alpha\left(\frac{d\alpha}{dr}\right)^2 \right]\\&+(k+\bar{k})\int^R_\epsilon\frac{dr}{r}\left(\tan\beta_1+\tan\beta_2\right)\\&-(k+\bar{k})\cos\alpha(R)(\pi+\phi_1(R)+\phi_2(R))\\&+(k+\bar{k})\cos\alpha(\epsilon)(\pi+\phi_1(\epsilon)+\phi_2(\epsilon))
\end{split}
\end{equation}
\begin{equation}
\label{eq:Fedge}
\begin{split}
F_{edge_i}=&\int^R_\epsilon dr\left[\frac{\gamma_i}{\cos\alpha\cos\beta_i}+\kappa_b \frac{\kappa_{g, i}^2+\kappa_{n, i}^2}{\cos\alpha\cos\beta_i}\right.\\&\left.-h_i\frac{\sin{\beta_i}}{\cos\alpha}\left(\kappa_2-\kappa_1\right)\right]\,,
\end{split} 
\end{equation}
\begin{equation}
\label{eq:constraint}
F_{\lambda_i}=\int^R_\epsilon dr \lambda_i\left(r\cos\alpha\frac{d\phi_i}{dr}-\tan{\beta_i}\right)\,.
\end{equation}
The index of the edges is $i=\{1,2\}$, $\epsilon$ is the cut-off length, which is given by $\epsilon\equiv-R\sin\alpha(R)$ for the catenoid.

\subsection{Euler-Lagrange equations}

The variations $\frac{\delta F}{\delta g_j}$, where $g_j=\{\alpha, \beta_1, \beta_2, \phi_1, \phi_2\}$, are equivalent to the Euler-Lagrange (EL) equations. For one independent and several dependent variables, the EL equations are given by~\cite{Arfken}

\begin{equation}
\label{eq:EL}
\frac{\partial f}{\partial g_j}=\frac{d}{dx}\frac{\partial f}{\partial g'_j}
\end{equation}
where $f=f(r, g_j, g'_j)$ is the free energy density given by $F=\int dr f(r, g_j, g'_j)$. Primes denote derivatives $\frac{dg_j}{dr}$. The set of EL Equations in Eq.~\eqref{eq:EL} contains ten coupled nonlinear first-order differential equations. Evaluating Eq.~\eqref{eq:EL}, these equations are calculated as
\begin{equation}
\label{eq:alphaderiv}
\begin{split}
u_1\equiv\frac{\partial f}{\partial\alpha'}=&\pi k r \cos\alpha\frac{d\alpha}{dr}+2\kappa_b\kappa_{n, 1}\cos\beta_1+2\kappa_b\kappa_{n, 2}\cos\beta_2\\&+h_1\sin\beta_1+h_2\sin\beta_2\,,
\end{split}
\end{equation}
\begin{equation}
\label{eq:betaderiv}
u_2\equiv\frac{\partial f}{\partial \beta_1'}=2\kappa_b\kappa_{g, 1}\,,\quad u_3\equiv\frac{\partial f}{\partial \beta_2'}=2\kappa_b\kappa_{g, 2}\,,
\end{equation}
\begin{equation}
\label{eq:phideriv}
u_4\equiv\frac{\partial f}{\partial \phi_1'}=\lambda_1r\cos\alpha\,,\quad u_5\equiv\frac{\partial f}{\partial \phi_2'}=\lambda_2r\cos\alpha\,,
\end{equation}
\begin{equation}
\label{eq:alpha}
\begin{split}
u_1' \cos^2\alpha&=\sigma \pi r\sin\alpha+\frac{k\pi\sin\alpha}{2}\left[\frac{2}{r}-\frac{\sin^2\alpha}{r}-r\cos^2\alpha\left(\frac{d\alpha}{dr}\right)^2\right]\\&+\sin\alpha\sec\beta_1 (\gamma_1+\kappa_b\kappa_{n, 1}^2-\kappa_b\kappa_{g, 1}^2)\\&+\sin\alpha\sec\beta_2 (\gamma_2+\kappa_b\kappa_{n, 2}^2-\kappa_b\kappa_{g, 2}^2)\\&+2\kappa_b\kappa_{n, 1}\sec{\beta_1}\left(\frac{\cos^2\alpha\sin^2\beta_1}{r}-\frac{\sin2\alpha}{2}\frac{d\alpha}{dr}\cos^2\beta_1\right)
\\&+2\kappa_b\kappa_{n, 2}\sec{\beta_2}\left(\frac{\cos^2\alpha\sin^2\beta_2}{r}-\frac{\sin2\alpha}{2}\frac{d\alpha}{dr}\cos^2\beta_2\right)\\&-\frac{1}{r}\left(h_1\sin\beta_1+h_2\sin\beta_2\right)\\&-\frac{\sin2\alpha}{2}(\lambda_1\tan\beta_1+\lambda_2\tan\beta_2)\,,
\end{split} 
\end{equation}
\begin{equation}
\label{eq:beta1}
\begin{split}
\cos\alpha\cos^2\beta_1 u_2'&=(k+\bar{k})\frac{\cos\alpha}{r}+\sin\beta_1(\gamma_1-2\kappa_b\kappa_{g, 1}^2)\\&+\kappa_b\sin\beta_1(\kappa_{n, 1}^2+\kappa_{g, 1}^2)+\frac{2 \kappa_b}{r}\kappa_{g, 1}\cos\alpha \\&+\left(2\kappa_b\kappa_{n, 1}\cos\beta_1\sin 2\beta_1-h_1\cos^3\beta_1\right)\\&\times\left(\frac{\sin\alpha}{r}-\cos\alpha\frac{d\alpha}{dr}\right)-\lambda_1\cos\alpha\,,
\end{split}
\end{equation}
\begin{equation}
\label{eq:beta2}
\begin{split}
\cos\alpha\cos^2\beta_2 u_3'&=(k+\bar{k})\frac{\cos\alpha}{r}+\sin\beta_2(\gamma_2-2\kappa_b\kappa_{g, 2}^2)\\&+\kappa_b\sin\beta_2(\kappa_{n, 2}^2+\kappa_{g, 2}^2)+\frac{2 \kappa_b}{r}\kappa_{g, 2}\cos\alpha \\&+\left(2\kappa_b\kappa_{n, 2}\cos\beta_2\sin 2\beta_2-h_2\cos^3\beta_2\right)\\&\times\left(\frac{\sin\alpha}{r}-\cos\alpha\frac{d\alpha}{dr}\right)-\lambda_2\cos\alpha\,,
\end{split}
\end{equation}
\begin{equation}
\label{eq:phi}
u_4'=u_5'=\frac{\sigma r}{\cos\alpha}+\frac{k}{2r}\sin\alpha\tan\alpha+\frac{k}{2} r\cos\alpha\left(\frac{d\alpha}{dr}\right)^2\,.
\end{equation}

In Eq.~\eqref{eq:alpha}, the two constraints $r\cos\alpha\frac{d\phi_1}{dr}=\tan\beta_i$ are used. Based on Eqs.~\eqref{eq:phideriv} and~\eqref{eq:phi}, we conclude that $\lambda\equiv\lambda_1=\lambda_2$. Then, the nine unknowns to be solved are $\alpha, \alpha', \beta_1, \beta_1', \beta_2, \beta_2', \phi_1, \phi_2, \lambda$.

Note that we neglect the contribution of the variations $\frac{\delta F}{\delta \phi_i}$ to the surface and edge deformations, as explained in the main text. Therefore, the structure in our approximation is governed solely by $\alpha, \alpha', \beta_1, \beta_1', \beta_2, \beta_2'$. In the case of the minimal surface, since $\alpha$ and $\alpha'$ are calculated analytically (see main text), the equations governing $\beta_1, \beta_1'$ and $\beta_2, \beta_2'$ are decoupled from each other.

\subsection{Boundary conditions}

The torque-free boundary conditions are given by Eqs.~\eqref{eq:alphaderiv},~\eqref{eq:betaderiv}, and~\eqref{eq:phideriv} being equal to zero~\cite{Arfken}. Sufficiently away from the rotation axis of the surface, in 2D membrane limit, $\alpha=0$, $\alpha'=0$, and $\alpha''=0$. Then, $\kappa_{n, i}=0$. Plugging in these results to Eq.~\eqref{eq:alpha}, we find that $\sin{\beta_i}=A_i/r$. This rather expected result applies to a straight line in polar coordinates, where the shortest distance between the origin and the straight line is $A_i$ (or, the in-plane protrusion amplitude). When $h_i=0$ as in the achiral limit, then the same relation still holds, and this time the torque-free condition $\frac{\partial f}{\partial \alpha'}=0$ is satisfied. The geodesic curvature $\kappa_{g, i}$ of a straight line vanishes in 2D, hence $\frac{\partial f}{\partial \beta_i'}=0$. In the 2D membrane limit $\lambda$ must be zero. This is best seen when a curve with the free energy given in Eq.~\eqref{eq:Fedge} is minimized 
on a plane. The resulting EL equations are independent of $\phi_i$, then $\lambda=0$. Hence $\frac{\partial f}{\partial \phi_i'}=0$, away from the rotation axis of the nonplanar surface.  Plugging in these results into Eqs.~\eqref{eq:beta1} and~\eqref{eq:beta2}, we find that $k+\bar{k}=\gamma_1 A_1=\gamma_2 A_2$. For a minimal surface, $k$ does not contribute to this equation.

At the cusp, ideally, the boundary conditions are $\alpha=-\pi/2$ and $\beta_i=0$ (again, we ignore $\frac{\delta F}{\delta \phi_i}$ dependence of the structure). In the case of the minimal surface, since $\epsilon\approx10^{-2} R$, we take $\alpha=0.95$ which nearly corresponds to a $70^o$ slope. This slope is steep enough as evidenced in Fig. 3(a) in the main text.